\documentclass{aa}
\usepackage{graphicx}
\usepackage{txfonts}
\usepackage{hyperref}
\usepackage[normalem]{ulem}
\hypersetup{colorlinks = true, citecolor = {blue}, urlcolor= {blue}}

\def\PGPU{$\varphi-$GPU }

\def\gapprox{\;\rlap{\lower 3.0pt                       
        \hbox{$\sim$}}\raise 2.5pt\hbox{$>$}\;}
\def\lapprox{\;\rlap{\lower 3.1pt                       
        \hbox{$\sim$}}\raise 2.7pt\hbox{$<$}\;}

\newcommand{\be}{ \begin{equation} }
\newcommand{\ee}{\end{equation}}

\newcommand{\ben}{\begin{enumerate}}
\newcommand{\een}{\end{enumerate}}

\renewenvironment{thebibliography}[1]{\begin{oldthebibliography}{#1}\setlength{\parskip}{0ex}\setlength{\itemsep}{0ex}}{\end{oldthebibliography}}

\usepackage{diagbox}
\usepackage{siunitx}
\newcommand{\orcid}[1]{\href{https://orcid.org/#1}{\protect\includegraphics[width=8pt]{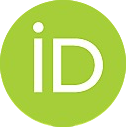}}}
\makeatletter
\renewcommand*\aa@pageof{, page \thepage{} of \pageref*{LastPage}}
\makeatother

\usepackage[dvipsnames]{xcolor}

\begin{document}
   
\title{Gravitational influence of the globular cluster NGC 7078 (M 15) flyby of the Oort cloud system}

\author{Maryna~Ishchenko
\inst{1, 2, 3}\orcid{0000-0002-6961-8170 }
\and
Peter~Berczik
\inst{2, 4, 1}\orcid{0000-0003-4176-152X}
}

\institute{Main Astronomical Observatory, National Academy of Sciences of Ukraine,
27 Akademika Zabolotnoho St, 03143 Kyiv, Ukraine           
\email{\href{mailto:marina@mao.kiev.ua}{marina@mao.kiev.ua}}
\and
Nicolaus Copernicus Astronomical Centre Polish Academy of Sciences, ul. Bartycka 18, 00-716 Warsaw, Poland
\and
Fesenkov Astrophysical Institute, Observatory 23, 050020 Almaty, Kazakhstan
\and
Konkoly Observatory, Research Centre for Astronomy and Earth Sciences, E\"otv\"os Lor\'and Research Network (ELKH), MTA Centre of Excellence, Konkoly Thege Mikl\'os \'ut 15-17, 1121 Budapest, Hungary
}
   
\date{Received xxx / Accepted xxx}
    
\abstract    
{It is crucial to understand the interaction between globular clusters (GCs) and the Oort cloud, as close flybys of such massive objects can significantly disturb the cloud's structure and redirect comets towards the inner Solar System. This increases the risk of impacts on Earth. Studying such events can teach us about the evolution and stability of the Solar System, as well as the effect of external gravitational forces on its dynamics over time.}
{In our study of the gravitational effects of the flyby of the NGC 7078 or M 15 GC on the Oort cloud, we focus on two types of approximation. First, we investigate the impact on the Sun's orbit during close passages, treating the GC as a point mass. At the second stage, we use a complete $N$-body system representation of the GC comprising over one million particles. The ultimate goal of the research is to quantify the number of particles stripped from the Oort cloud, and to understand the conditions under which this occurs.}
{We carried out a dynamical study of the gravitational interaction between Oort cloud particles and galactic GCs within the time-varying galactic external potential. Initially, the GCs are represented as point masses orbiting the Galaxy alongside the Sun and the Oort cloud system. This study was also extended to include the case of NGC 7078, for which full $N$-body long-term dynamical modelling of the GC itself was used.}
{Our study reveals significant variations in the impact of NGC 7078 on the Oort cloud, depending on whether it is modelled as a point mass or a complete $N$-body system. The $N$-body system results in much greater stripping of Oort cloud particles, with over 52\% stripped during a close pass, compared to a few percent in the point mass model for a flyby at a large distance (>200 pc) and 36\% for a closer 10 pc point mass flyby. The N-body system also causes substantial expansion, with particles spreading over 50 pc from the Sun within 30 Myr after the GC's crossing. This creates a twisted and flattened cloud structure with extended outer tails. These stripped cloud particles (more than 10\%) spread across the galaxy, reaching distances of up to 16 kpc from the Sun. These differences emphasise the importance of using detailed N-body simulations to accurately evaluate the gravitational influence of GCs on the Oort cloud and shed light on the varying effects of simple versus complex system representations.}
{}



\titlerunning{NGC 7078 flyby onto the Oort cloud system}
\authorrunning{M.~Ishchenko et al.}
\maketitle

\section{Introduction}\label{sec:Intr}

The current success of the Gaia mission enables us to predict the dynamic evolution of our Solar System within the broader Galactic framework with a much higher degree of precision than before. Implementing the time-variable potential can increase the reliability of the results obtained, such as the ones from orbital integration over long cosmological timescales. One such tool is the IllustrisTNG-100 cosmological database, which provides a Milky Way (MW)-like combination of gravitational potential and disc.

In our previous study \cite{Ishchenko2024-4} (hereafter \hyperlink{I23}{\color{blue}{Paper~IV}}), we examined the statistical interaction between globular clusters (GCs) and the Solar System over a six-billion-year period. Our aim was to estimate the gravitational influence of GCs' flybys on the Oort cloud system. To model the Galactic orbital dynamics of each cluster realistically, we used our own high-order parallel $N$-body GPU code. To identify close passages of clusters with the Solar System, we adopted a simple relative distance criterion of below 200 pc.

Here, we briefly summarise our main results obtained in Paper IV. We identified 35 GCs that could potentially experience close encounters with the Sun across all selected gravitational potentials throughout the Sun's entire lifetime. Our selection of different IllustrisTNG-100 MW halo-disc potentials was primarily based on reproducing the global mass and size parameters of the current MW system. As well as these current parameters (namely two masses and three scale sizes; see Sect. 2.2 in \cite{Ishchenko2023a}), we also considered the time evolution of the circular velocity near the vicinity of the Sun (Sect. 2.2, Appendix A in \cite{Ishchenko2023a}. We concluded that, on average, there is a $\sim$15\% probability of a close encounter over six billion years for a GC at a relative distance of 50 parsecs from the Solar System, across all six gravitational potentials. At a distance of 100 parsecs, this probability increases to $\sim$35\%. The GCs BH 140, UKS 1, and Djorg 1 have mean minimum passage distances to the Sun of 9, 19, and 17 pc, respectively.

Flyby GCs can significantly impact the Solar System by disturbing the Oort cloud. This could send cometary nuclei inward, increasing the likelihood of collisions with Earth and potentially altering the planet's environment. During close encounters, their gravitational interactions could slightly alter the orbits of planets, adding complexity to the dynamics of the Solar System. These rare yet significant events highlight the interconnectedness of our cosmic environment and emphasise the importance of studying these scenarios to understand how they have shaped the history of the Solar System and the evolution of Earth. The gravitational effects of a GC can affect long-term stability and potentially trigger environmental changes on planets that could have far-reaching consequences. Analysing the probabilities and impacts of flybys enhances our understanding of the role that GCs play in shaping the dynamics of the Solar System.

This study aims to analyse the gravitational influence of a flyby of individual GCs on the Oort cloud in more detail. The analysis is carried out for two types of gravitational interaction. The first type involves analysing the gravitational influence during a close passage of the Sun's orbit. In this case, the GC is represented as a single point mass interacting with the Oort cloud system at different relative distances. In order to reduce the large initial dynamical parameter space, and based on the best local representation of the solar neighbourhood's global dynamical evolution (value of the circular velocity), we selected only one time-variable potential (TVP) for further modelling: {\tt 411321} TVP.

In the second stage, we analysed the gravitational influence on the Oort cloud system from the GC presented as a full $N$-body system. For this purpose, we selected the most interesting close passage case from the previous stage. Here, the individual GC was initially generated with more than one million particles and with a mass of around $\sim$10$^6$ M$\odot$. The modelling was performed on a star-by-star basis, taking into account also the updated stellar evolution scheme for the GC stellar particles. From this, we estimated the number of detached particles from the Oort cloud due to a nearby flyby $N$-body GC. Summing up, we provide information about the stripped particles from the Oort cloud due to a variety of conditions. 

The paper is organised as follows. In Sect. \ref{sec:rate} we analyse the most probable individual GC close passage near the Sun in {\tt 411321} TVP external potential. In Sect. \ref{sec:ini-oort} we describe the initial conditions and the formation process of the Oort cloud system. In Sect. \ref{sec:point-oort} we present the gravitational influence on the Oort cloud system due to the flyby NGC 7078 as a point mass. Sect. \ref{sec:7078-full-oort} is dedicated to the gravitational influence on the Oort cloud system due to the flyby of NGC 7078 as a full $N$-body system. We summarise and discuss our results in Sect. \ref{sec:disc}. 

\section{Analysis of the most probable close passage GCs near the Solar System}\label{sec:rate}

\begin{figure}[htbp!]
\centering
\includegraphics[width=0.92\linewidth]{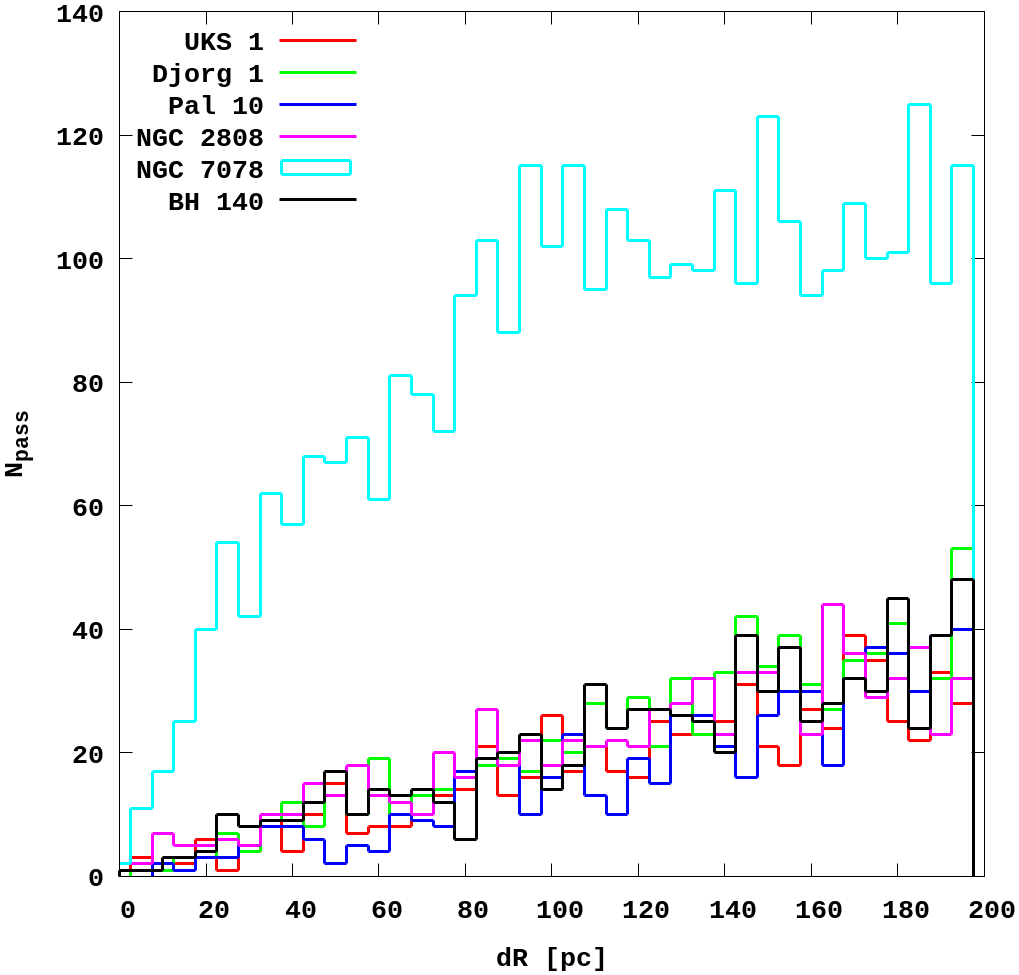}
\caption{Average cumulative number of passages between GCs and the Solar System is shown as a function of the relative distance, $dR$. Colours represent individual GCs.}
\label{fig:dr_fit}
\end{figure}
\begin{figure}[htbp!]
\centering
\includegraphics[width=0.99\linewidth]{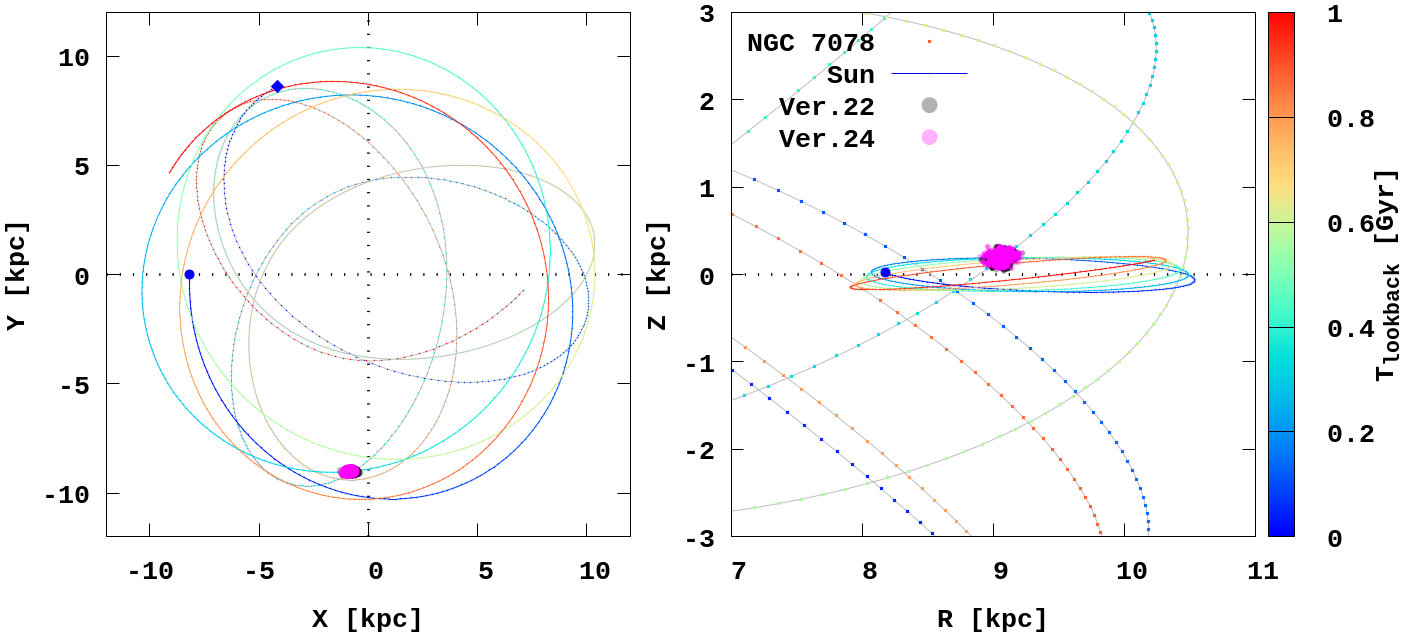}
\caption{Orbital solar evolution (solid coloured line) with close passes of NGC 7078 (coloured dots on grey line) during the 1 Gyr integration in time for {\tt 411321} TVP. The coloured circles represent the collisions based on {\tt Ver.24}, with magenta and black representing the {\tt Ver.22} catalogue. The blue circle and rhombus represent the current positions of the Sun and NGC 7078, respectively.}
\label{fig:sun_orb_7078-1-gyr}
\end{figure}

In this section, we focus on estimating close passages (interactions) of the GCs BH 140, Djorg 1, UKS 1, Palomar 10, NGC 7078, and NGC 2808. These GCs were selected because they have already demonstrated close passages in our set of TVPs, as is shown in Table A.1 in \hyperlink{I23}{\color{blue}{Paper~IV}}. For the current research, we used up-to-date 6D GC information for positions, proper motions, radial velocities, and heliocentric distances. This information was collected from the website\footnote{\label{note1} 6D GC information \url{https://people.smp.uq.edu.au/HolgerBaumgardt/globular/orbits_table.txt}} of \cite{VasBaum2021}. We re-integrated all of these six GCs, using the currently available catalogue data from the catalogue, together with the Sun as a point mass in a {\tt 411321} TVP MW-like potential using the same as in previous research: \PGPU\footnote{$N$-body code \PGPU: \\~\url{ https://github.com/berczik/phi-GPU-mole}} up to 5 Gyr of lookback time.

\begin{figure}[htbp]
\centering
\includegraphics[width=0.95\linewidth]{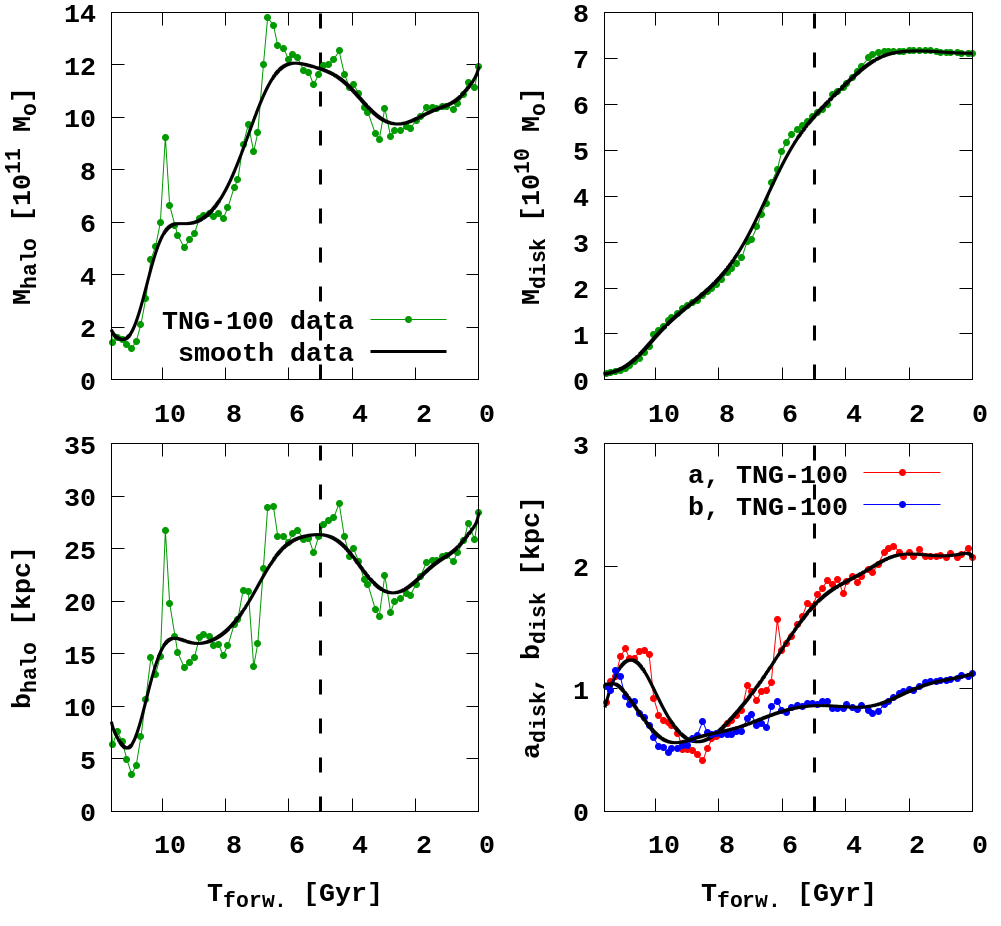}
\caption{Evolution of the halo and disc masses, and their characteristic scales for {\tt 411321} TVP. Green, red, and blue lines with dots show the parameters recovered from the IllustrisTNG-100 data. Solid black lines correspond to the values after the interpolation and smoothing with a time step of 1 Myr, which was used in the orbital integration. The dashed lines represent the 5 Gyr duration.}
\label{fig:MW-TNG}
\end{figure}

Fig.~\ref{fig:MW-TNG} shows the evolution of halo and disc masses, and their characteristic scales for the external potential {\tt 411321}, which was selected from the IllustrisTNG-100 cosmological database. To obtain the spatial scales of the discs and dark matter haloes, we decomposed the mass distribution using the Miyamoto-Nagai (MN)~$\Phi_{\rm d} (R,z)$ \citep{Miyamoto1975} and Navarro–Frenk–White (NFW) $\Phi_{\rm h} (R,z)$ \citep{NFW1997} potentials:
\begin{equation}
\begin{split}
\Phi_{\rm tot} &= \Phi_{\rm d} (R,z) + \Phi_{\rm h} (R,z) = \\
&= - \frac{GM_{\rm d}}{\sqrt{R^{2}+\Bigl(a_{\rm d}+\sqrt{z^{2}+b^{2}_{\rm d}}\Bigr)^{2}}} - 
\frac{GM_{\rm h}\cdot{\rm ln}\Bigr(1+\frac{\sqrt{R^{2}+z^{2}}}{b_{\rm h}}\Bigl)}{\sqrt{R^{2}+z^{2}}},
\end{split}
\end{equation}
where 
$R=\sqrt{x^{2}+y^{2}}$ is the planar galactocentric radius, 
$z$ is the distance above the plane of the disc, 
$G$ is the gravitational constant, 
$a_{\rm d}$ is the disc scale length, 
$b_{\rm d,h}$ are the disc and halo scale heights, respectively, and 
$M_{\rm d}$ and 
$M_{\rm h}=4\pi\rho_{0}b^{3}_{\rm h}$ ($\rho_{0}$ is the central mass density of the halo) are the masses of the disc and halo, respectively.

More details about the selected {\tt 411321} TVP can be found in \cite{Ishchenko2023a} and \cite{Ishchenko2024-4}, and also on our website.\footnote{MW-like TVP: \\~\url{https://sites.google.com/view/mw-type-sub-halos-from-illustr/TNG-MWl?authuser=0}} For event detection of the closed passages of the GCs, $N_{\rm pass}$, we applied the same simple criterion: the separation, $dR$, between the GC and the Solar System should be less than 200~pc.

To achieve statistically significant results, we conducted 10 000 independent simulations for each GCs in which the initial proper motions ($\mu_{\rm \alpha}$, $\mu_{\rm \delta}$), radial velocity, and heliocentric distance of the GCs were varied according to the normal distribution of $\pm1\sigma$ with their respective measurement errors, listed in the referenced catalogue based on the year 2024. We refer to this version of the catalogue as {\tt Ver.24} (please refer to App.~\ref{app:catal-recalc} for more details on the statistical probability of close passages based on two versions of the same catalogue {\tt Ver.24} and {\tt Ver.22}). 

In Fig.~\ref{fig:dr_fit} we present the histogram of close pass interactions (crossings) between the orbits of GCs and the Solar System -- $N_{\rm pass}$ -- as a function of the relative distance, $dR$, in the {\tt 411321} TVP. These statistics represent the sum of all individual randomisations for a total of 10 000 runs. Each GC is presented with different colours. As can be seen, five GCs, except for NGC 7078, exhibit a fairly similar `linear' behaviour, with very similar $N_{\rm pass}$ values for each GC. For example, at $dR = 50$ pc there are approximately 10 events, and at $dR = 100$ pc there are approximately $\sim$18--20 events. However, the behaviour of NGC 7078 is significantly different. Firstly, the $N_{\rm pass}$ values are significantly higher: for the same $dR$s we have, correspondingly, $\sim$55 and $\sim$85 events, i.e. a factor of $\sim$4--5 times more. Furthermore, the dependency of $N_{\rm pass}$ versus $dR$ is a growing function, but only up to $\sim$100 pc. Above this limit, there is almost a plateau of around 100 passages in total. Furthermore, NGC 7078 has a relatively high probability of close passages: around $\sim$33\% for our $10k$ random seeds and for both versions of the catalogue (see Table \ref{tab:compar}). 

\begin{table}[tbp]
\setlength{\tabcolsep}{4pt}
\centering
\caption{Comparison of the close passages based on two versions of the catalogue in {\tt 411321} TVP.}
\label{tab:compar}
\begin{tabular}{lcccc}
\hline
\hline 
GC & \multicolumn{2}{c}{{\tt Ver.22}} & \multicolumn{2}{c}{{\tt Ver.24}} \\
   & Prob., \% & $dR_{min}$, pc & Prob., \% & $dR_{min}$, pc \\
\hline
\hline
BH 140 & 8.0 & 3.3 &  9.4 & 6.9  \\
Djorg 1 & 8.5 & 20 &  8.3 & 8.8  \\
UKS 1 & 7.8 & 3.9 &  7.0 & 9.7  \\
NGC 7078 & 34.6 & 8.2 & 33.6 & 2.0 \\
NGC 2808 & 8.0 & 33 & 7.3 & 4.3 \\
Palomar 10 & 5.6 & 11 & 7.7 & 9.3  \\
\hline 
\end{tabular}
\vspace{6pt}
\end{table}

Additionally, in Fig. \ref{fig:dr_fit2} we show the relative velocities of the closest passages of the GCs and the corresponding time bins for our set of GCs. For the most probable close passage candidate NGC 7078 we identify the time of the `special' (prominent) passage (with a $\sim$30\% probability across all random seeds) at $\sim$332 Myr lookback time with a relative velocity of $\sim$100--150 km s$^{-1}$.

From all our orbit reconstruction runs, we can conclude that, at the time of close passages between the Sun and NGC 7078, our initial data randomisation, and taking into account the \textit{Gaia} DR3 measurement errors, the possible detected minimum $dR$ was 2 pc (in only two random seeds). Clearly, in a real galactic dynamical system, such a small flyby separation is highly improbable ($\sim$0.02\%). Based on this, we propose a working model of a minimum passage distance of 10 pc, which has been observed in dozens of random seeds (i.e. with a probability of around $\sim$0.1\%). Our assumption of a minimum distance of 10 pc ($dR_{min}$) overlaps with the minimum distance ($dR$) for the two versions of the GC catalogue {\tt Ver.24} and {\tt Ver.22} (see Table \ref{tab:compar}). We refer to this selected close passage as the `basic' model. 

Here, in Fig.~\ref{fig:sun_orb_7078-1-gyr} we demonstrate the orbital evolution of the NGC 7078, which is our main candidate with a strong gravitational influence on the Oort cloud system. This plot shows the simultaneous evolution of the Sun and the NGC 7078 orbits simultaneously up to 1 Gyr of lookback time. The full orbital evolution up to 5 Gyr is presented in Fig. \ref{fig:sun_orb_7078}. 

In summary, we have selected NGC 7078 (M 15) for our future analysis of the gravitational impact on the Oort cloud system. This GC is also one of the oldest known GCs with an estimated age of around 12 billion years \citep{OMalley}. It is also one of the most densely packed GCs in the MW. The current half-mass radius is 3.66 pc, and its estimated mass is 5.18$\times$10$^5$ M$_{\rm \odot}$ \citep{Baumgardt2021}. Its core has already undergone a `core collapse', resulting in a steep central density cusp. This dense core may even harbour an intermediate-mass black hole \citep{Gerssen2003}. 

\section{Initial conditions of the Oort cloud system} 
\label{sec:ini-oort}

To investigate the gravitational impact of the flyby of NGC 7078 on particles in the Oort cloud, we generated 50 000 massless cloud particles. The initial conditions are the same as we discussed in \hyperlink{I23}{\color{blue}{Paper~IV}}. Here, we only briefly reiterate the main parameters of the cloud. 

For the Oort cloud, we used the Dehnen profile \citep{dehnen_family_1993} as a well-established initial equilibrium particle distribution \citep[see Figure~9 in][]{Portegies2021b} with the power slope $\gamma = 2$: 

\begin{equation}\label{eq:rho_D}
    \rho_{\rm Oort}(r) = \frac{(3-\gamma) \cdot M_\mathrm{Oort}}{4\pi} \frac{a_\mathrm{Oort}}{r^\gamma (r+a_\mathrm{Oort})^{4-\gamma} },
\end{equation}
where M$_{\rm Oort}$ is the total mass of the Oort cloud model (in our case $0.01\rm\;M_{\odot}$) and $a_\mathrm{Oort}$ is the Oort cloud scaling radius (which we set as $10^{5}$~au). 
 
\begin{figure}[htbp!]
\centering
\includegraphics[width=0.99\linewidth]{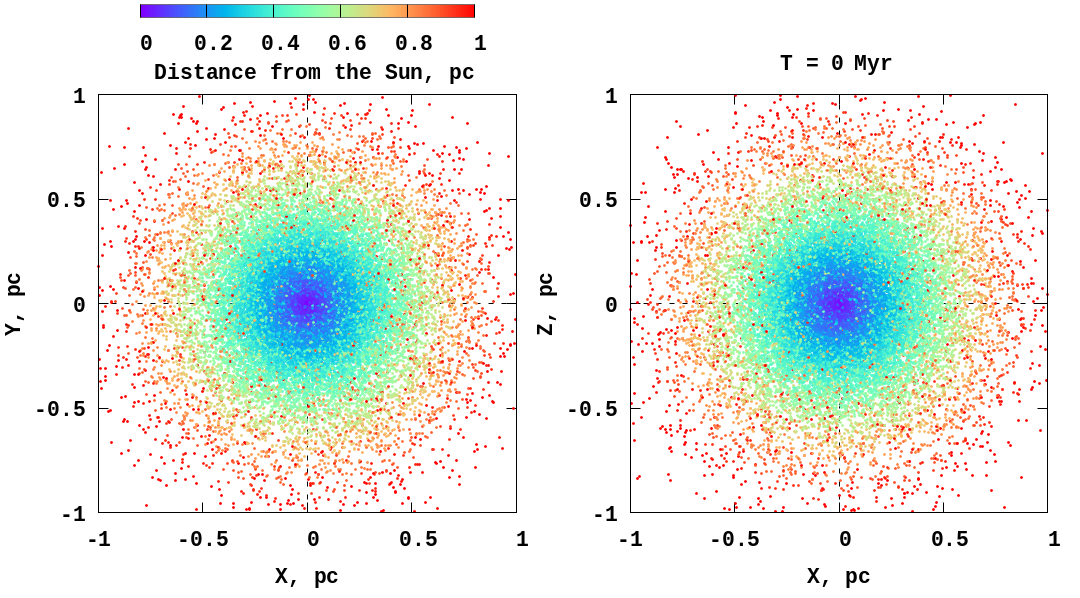}
\caption{Initial particle configuration for the Oort cloud system with N = 50$k$.}
\label{fig:init-oort}
\end{figure}

For actual data generation, we used the popular {\tt AGAMA} library \citep{agama2019}. The inner and outer cut-off radii for the particle distribution model were set at 100~au ($\sim4.8\times10^{-4}$~pc) and 1~pc ($\sim2\times10^{5}$~au). For the outer radius truncation, we simply used the Sun Hill radius definition in the Galactic potential, which is around $\sim$1 pc, as in \citealt{Portegies2021a}. The initial equilibrium distribution of particles generated around the Sun only takes into account the Sun's gravitational field. The resulting initial particle distribution around the Sun is presented in Fig.~\ref{fig:init-oort}. 

\section{Point mass gravitational influence of the NGC 7078 to the Oort system} \label{sec:point-oort}

Next, we analysed the influence of the point mass NGC 7078 on the distribution of Oort cloud particles around the Sun during dynamical orbital integration in the external TVP MW-like potential {\tt 411321}.  As was noted above, we present the NGC 7078 as a point mass object with a current observed mass of 5.18$\times10^{5}$ M$\odot$. Following \cite{Portegies2021b}, we neglected the self-gravity of the Oort cloud particles in our investigation and calculated the gravitational interaction between the point mass Sun and the point mass NGC 7078, as well as their mutual gravitational influence on the Oort cloud particles. In other words, the Oort system was treated as massless particles.

\begin{figure}[htbp!]
\centering
\includegraphics[width=0.99\linewidth]{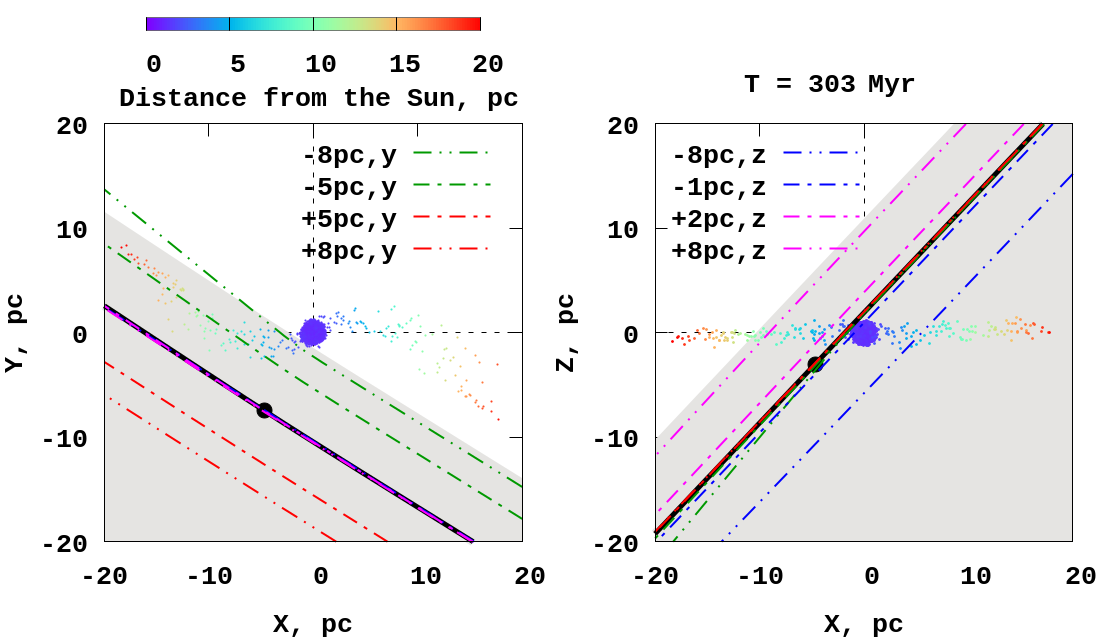}
\caption{Orbital trajectory of NGC 7078 near the Oort cloud system, centred on the Sun's position. The filled black circle indicates the cluster's position at the time of its closest approach, 332 Myr years ago.}
\label{fig:ini-pass}
\end{figure}

Fig. \ref{fig:ini-pass} shows the dynamical distribution of the Oort system centred on the Sun (i.e. with the Sun located at \textit{X-Y-Z} = 0) after integration up to a lookback time of 303 Myr. We can see that a small number of particles (0.1–0.2\%) leave the cloud due to their orbital motion in the time-variable external potential, primarily under the influence of solar gravity. However, the Oort cloud itself still has a fairly spherical shape (violet dots).

Next, we added the NGC 7078 point mass trajectories to this plot. The grey zone shows all possible NGC 7078 trajectories resulting from variations in the initial position and velocity due to random seeds (the influence of measurement errors in the initial conditions on the orbital trajectory, as is described in App. \ref{app:catal-recalc}).

\begin{figure}[htbp!]
\centering
\includegraphics[width=0.99\linewidth]{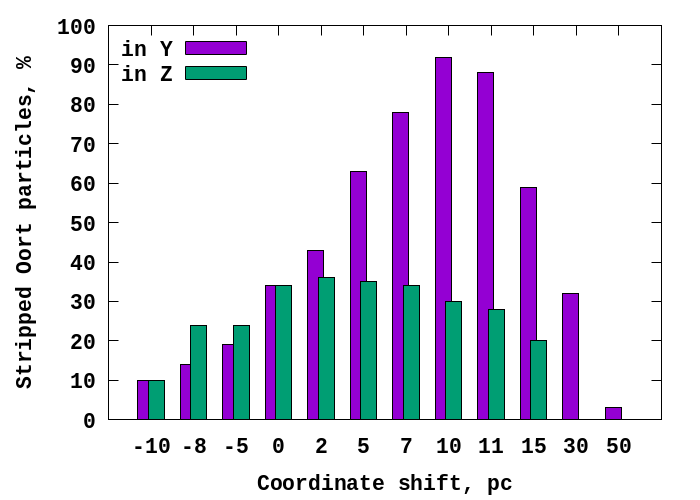}
\caption{Percentage of stripped Oort particles due to the NGC 7078 co-ordinate shift in the $Y$ and $Z$ co-ordinates separately. Our basic model with the $dR$ = 10 pc on the plot is marked as `0'.}
\label{fig:shift}
\end{figure}

In the figure, we also illustrate the basic case of the NGC 7078 orbit as a bold black line (see Sect. \ref{sec:rate}). This orbit has a relative distance of 10 pc at its closest approach to the Sun. We detected this cluster flyby near the Sun point at 332 Myr ago. Using this trajectory, we calculated how many particles from the Oort cloud were stripped by the passing NGC 7078. Under these conditions, we estimate that 36\% of particles were lost, as is shown in Fig. \ref{fig:shift}, in which our basic close passage is marked as `0' on the \textit{X} axis. 

Next, we varied the NGC 7078 Galactic co-ordinates \textit{Y} and \textit{Z} before 10 Myr of the moment of close passage, bringing the passage closer or farther away from the Sun. The co-ordinates were varied within the grey zone, i.e. within the possible trajectories of the cluster. We present some of the new shifted trajectories in \textit{Y-Z} as dashed lines in green and red for \textit{Y}-co-ordinate shift and in magenta and blue for \textit{Z}-co-ordinate shift (see Fig. \ref{fig:ini-pass}). It should be noted that the minimum closest passage can be at a distance of 2 pc from Oort, resulting in a slight deviation of the cluster's orbit itself (see the dash-dotted green and dotted blue lines).

We analyse in detail the amount of stripped particles from the Oort cloud resulting from the shifted NGC 7078 trajectories. The results are presented in Fig. \ref{fig:shift}. In the case of a shift along the \textit{Y} co-ordinate, the most destructive case for the Oort system is the passage of the NGC 7078 at +10 pc compared to the basic model, with the following values: $X_{shift}$ = 0, $Y_{shift}$ = +10, and $Z_{shift}$ = 0. In this case, 90\% of the cloud particles are lost (violet boxes, marked as `10 pc' on the figure abscissa). In the case of a shift along the \textit{Z}-co-ordinate, a maximum of 38\% of particles are lost at a relative distance of +2 pc from the basic model (green boxes, marked as `2 pc' on the figure abscissa), with the following values:$X_{shift}$ = 0, $Y_{shift}$ = 0, and $Z_{shift}$ = +2.

The high percentage of stripped Oort cloud particles presented above could be the result of our strictly point mass approximation of NGC 7078 as a whole. In the case of a smoother cluster mass (for example, a Plummer-like mass), we can expect a lower percentage of stripped particles. A more detailed discussion of gravitational stripping by a flyby cluster as a point mass system is presented in Sect. \ref{sec:disc}.   

\begin{figure}[htb!]
\centering
\includegraphics[width=0.95\linewidth]{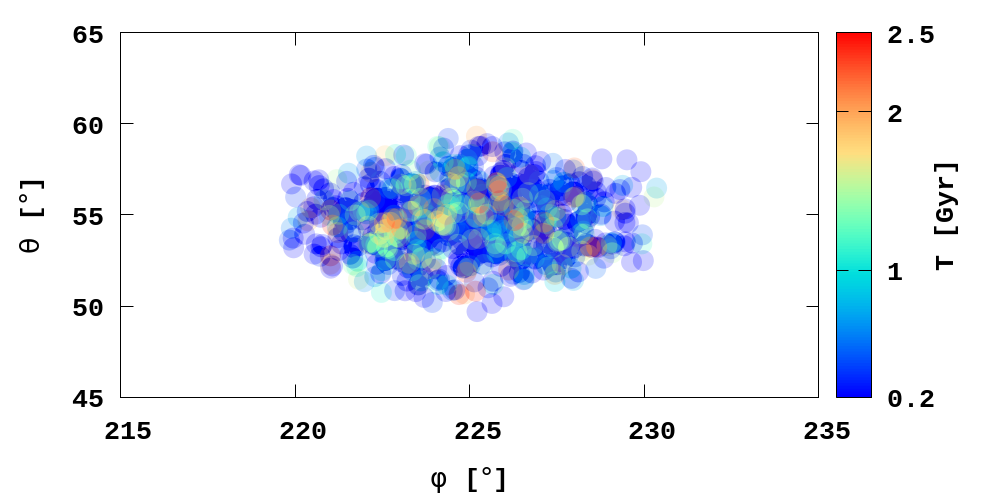}
\includegraphics[width=0.95\linewidth]{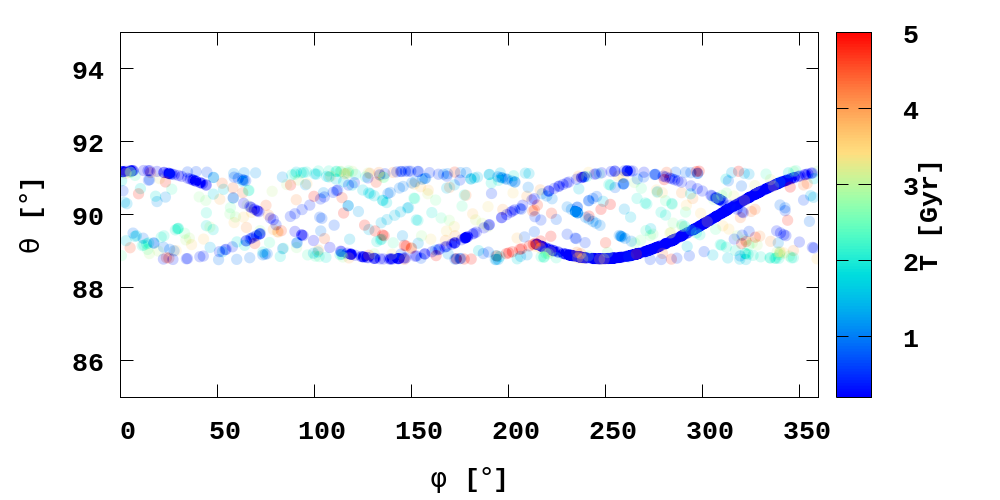}
\caption{Distribution of spherical angles of cross-passing events of Oort particles with a radius of 1 AU penetrating around the Sun. The upper panel shows the penetration of Oort particles when the closest relative distance between M 15 and the Sun is 10 pc. The bottom panel shows M 15 randomisation when the closest relative distance is more than 200 pc. The polar angle $\theta$ = 90$^{\circ}$ corresponds to the Galactic equator.}
\label{fig:ugol}
\end{figure}

To illustrate the penetration of particles into the inner Solar System (i.e. within 1 AU), we present the spherical co-ordinate angles of the penetration points of Oort cloud particles within a 1 AU radius sphere around the Sun in the upper panel of Fig.\ref{fig:ugol}. As can be seen in the figure, the most active phase of Oort particle penetration into the inner Solar System begins around the time of M 15's closest passage (around 300 Myr) and ends completely before the integration time reaches 2.5 Gyr. As can be seen in the plot, the most intense particle crossing occurs within the time interval of $\sim$300--500 Myr lookback time. There is strong evidence of a similar extra bombardment of the inner Solar System at the end of the Paleozoic era, which was also recently reported in the paper \cite{Mazrouei2019}. It is also worth noting the very narrow range of particles in this sphere, which may indicate penetration as a narrow stream of particles into the inner Solar System. The total number of Oort particles that crossed the 1 AU radius sphere during this period was $\approx$2\% of the initial number of particles in the cloud.

On the other hand, Fig. \ref{fig:ugol} (bottom panel) shows a completely different pattern of penetration angles in the case of M15 with different initial positions, when the closest approach to the Sun was >200 pc. In this case, the total number of penetrations to the 1 AU radius is similar ($\approx$2\%), but they are distributed over a 5 Gyr integration time. Also, we observe a wide range of azimuthal angles (360$^{\circ}$) and a narrow range of polar angles (a few degrees around 90$^{\circ}$).

\section{Star-by-star gravitational influence of the NGC 7078 on the Oort system} 
\label{sec:7078-full-oort}

\subsection{Initial conditions of NGC 7078 and the cluster integration} 
\label{subsec:init-cond-7078}

For the dynamical orbital integration of NGC 7078, we employed the same high-order, parallel $N$-body \PGPU code that we previously described, but with recently updated stellar evolution routines for the star particles \citep{Kamlah2022, Kamlah2022MNRAS}. We used the newly updated {\tt MCLUSTER} code \citep{Kamlah2022MNRAS, Kuepper2011} with the following parameters to generate the initial $N$-body model for NGC 7078. The individual stellar masses were based on the Kroupa initial mass function \citep{Kroupa2001}, with lower and upper mass limits set at 0.08--100~$\rm M_{\odot}$. To ensure the long-term survival of the NGC 7078 cluster, we adopted a fairly concentrated King model (W$_0$ = 9) for the initial distribution of particles, with a half-mass radius of 2 pc. We also adopted an individual metallicity of -2.28 [Fe/H] for this object (\citep{Bland-Hawthorn2016, Husser2020}. Within the code, we used the classical solar metallicity value $Z_{\odot} = 0.02$ for the scaling of our stellar evolution model \citep{Grevesse1998}. For the current cluster metallicity, we obtain the value $Z_m = 1.05\times10^{-4}$ in absolute units.

Additionally, we define the dynamical mass of each particle as ten times the mass of the star that it represents. Consequently, each particle in the simulation corresponds to ten similar stars, and stellar counts are scaled by a factor of 10 to obtain realistic values. This approach allows us to run a much larger number of fitting models to estimate the initial mass of NGC 7078 by comparing the results of the fitting runs with the cluster's present-day mass. Further information on the initial conditions and integration procedure can be found in \cite{Ishchenko2024mass-loss}. 

We begin the dynamical simulation of NGC 7078 at a lookback time of eight billion years and continue to the present day. The co-ordinates and corresponding velocity components in the galactocentric Cartesian reference frame (GCRF) at eight billion years ago were taken from \cite{Ishchenko2023a}. Table \ref{tab:init-param-7078} summarises our best-fitting initial conditions for NGC 7078.

\begin{table}[tbp]
\setlength{\tabcolsep}{4pt}
\centering
\caption{Initial kinematics and physical properties at eight billion years ago for NGC 7078 in {\tt 411321} TVP external potential.}
\label{tab:init-param-7078}
\begin{tabular}{lcc}
\hline
\hline 
Parameters & Unit & Value \\
\hline
\hline
   & $X$, kpc & -10.4 \\
Position in GCRF   & $Y$, kpc & -0.56 \\
  & $Z$, kpc & 5.9 \\
  & $V_X$, km s$^{-1}$ & -322  \\
Velocities in GCRF & $V_Y$, km s$^{-1}$ & 1.1  \\
  & $V_Z$, km s$^{-1}$ &  0.36 \\
Mass    & M$_{\rm \odot}$ & 9.2$\times$10$^{5}$  \\
Number of stars    &   & 1 603 165  \\
Radius of half-mass    & pc & 2.0  \\
Concentration (King profile)  & W$_0$ & 9.0 \\
\hline 
\end{tabular}
\vspace{6pt}
\end{table}

\subsection{Dynamical evolution of NGC 7078} 
\label{subsec:evol-7078}

\begin{figure}[htbp!] 
\centering
\includegraphics[width=0.99\linewidth]{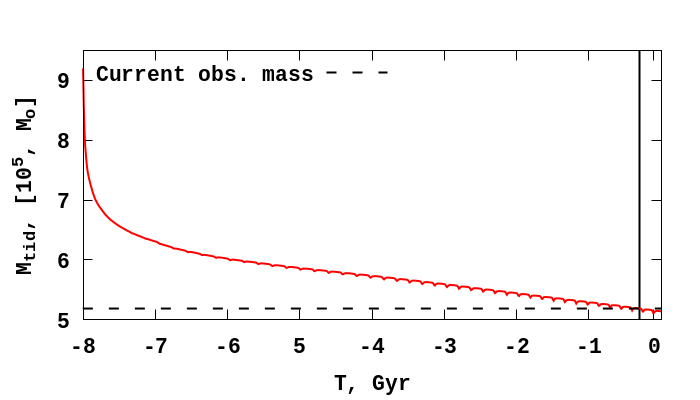}
\caption{Evolution of tidal mass as a function of time. The dashed line represents the currently observed mass. The vertical black line represents the time at which the GC passes close to the Sun.}
\label{fig:mass-loss-7078}
\end{figure}

Here, we briefly discuss the mass loss of NGC 7078, as determined by our $N$-body numerical integration. We varied the initial mass of the cluster, the half-mass radius, and the King concentration parameters (see Table \ref{tab:init-param-7078}) to determine the optimal conditions under which we would obtain the observed cluster today, with a mass difference of generally within $\pm5\%$ , after eight billion years of dynamical evolution (see Fig. \ref{fig:mass-loss-7078}).

\begin{figure*}[htbp!] 
\centering
\includegraphics[width=0.79\linewidth]{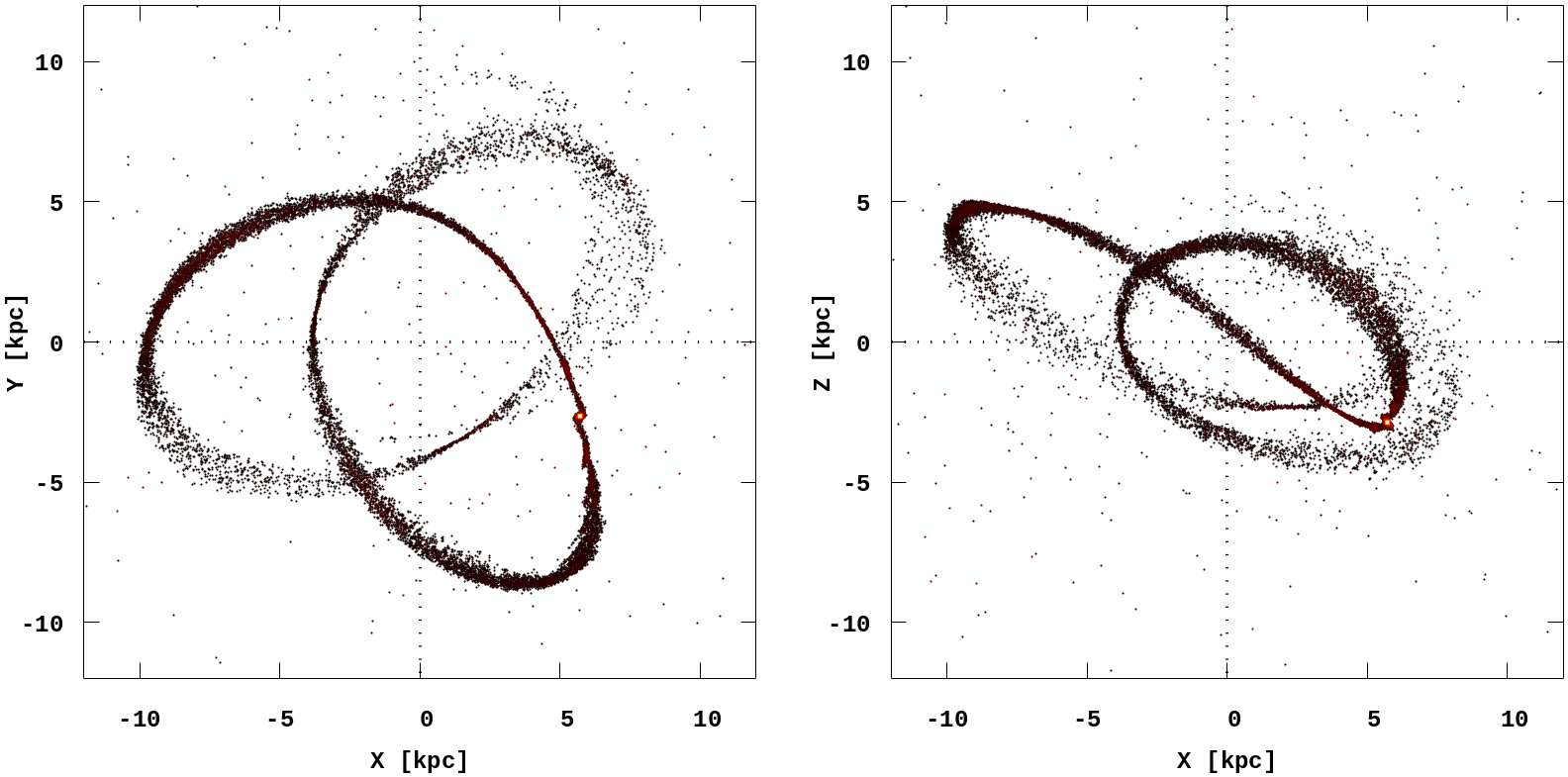}
\caption{Distribution of stars in NGC 7078 at present day in {\tt 411321} TVP. The orbital and stellar evolution is shown in two projections -- ($X$-$Y$) and ($X$-$Z$).}
\label{fig:def-den-7078}
\end{figure*}

During the first $\sim$100--150 Myr, around 20-25\% of the cluster's initial mass is lost, primarily due to stellar evolution processes. After this period, the main cause of cluster mass loss is tidal interaction with the external potential of the MW. Consequently, the cluster retains approximately $\sim$55\% of its original stellar mass to this day, almost at the moment of closest passage near the Sun (vertical black line in Fig. \ref{fig:mass-loss-7078}).

In Fig. \ref{fig:def-den-7078} we demonstrate the distribution of all stars on orbit after eight billion years of evolution (i.e. the present day). 
The core of NGC 7078 still contains a very compact stellar subsystem within the current half-mass radius of approximately $\sim$4 pc. Interestingly, more than 15\% of the cluster's stars are located in the trailing and leading tails (beyond 2$\times$r$_{tid}$).

\subsection{Co-evolution of NGC 7078 and the Oort cloud system} 
\label{subsec:evol-7078+oort}

After we had found the appropriate initial conditions for the dynamical model of NGC 7078, we added the Sun and Oort cloud system to the global simulation. To generate the internal structure of the Oort cloud system, we used the same procedure that we described in Sect. \ref{sec:point-oort}. We combined the Oort cloud model snapshot with the snapshot of the Sun and the cluster of stars to create one global particle snapshot at a lookback time of 320 Myr (just 10 Myr before the predicted closest approach). The Oort particles were added to the system as massless particles. Gravity interactions between the Oort cloud particles were ignored. Each cluster stellar particle individually affects the Oort cloud system via gravity. Naturally, the interactions between all the stars, the Oort particles, and the Sun particles were also taken into account. Of course, during the simulation, we based our calculations on selected numerical realisations (randomisations) of both the NGC 7078 and the Oort cloud system. An extended study of the influence of individual randomisations of particle distributions inside the M 15 and Oort clouds is beyond the scope of this paper, but will be addressed in future work on this topic.

\begin{figure}[htbp!] 
\centering
\includegraphics[width=0.79\linewidth]{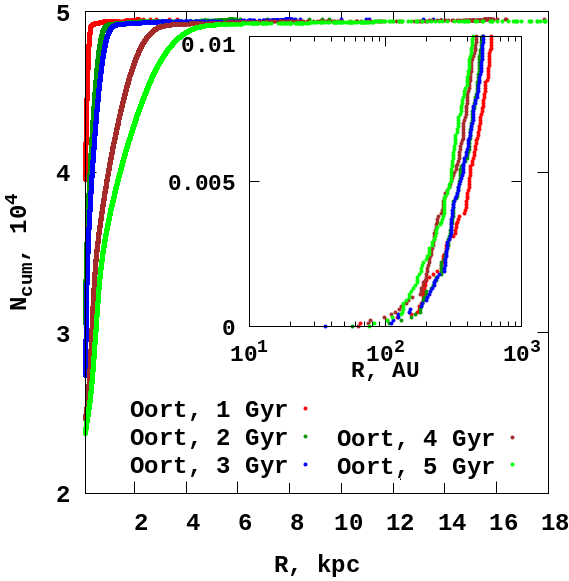}
\caption{Oort particle distribution due to flybys of NGC 7078 as a function of relative distance from the Sun at different times. The inner Oort particle distribution is presented in the embedded plot.}
\label{fig:dr-oort}
\end{figure}

In Fig. \ref{fig:dr-oort} we show the distribution of stripped Oort particles (at a distance from the Sun of more than 100 pc) due to the gravitational influence of the flyby of NGC 7078, which occurred 332 million years ago, at different points in the future following selected close passes. As can be seen, the stripped particles are distributed throughout the entire galaxy, at distances of up to 16 kpc from the Sun. The number of such particles is significant, at around 50\%. There are a few thousand of them at distances greater than 10 kpc.

Such particles from the Oort cloud can manifest as high-velocity intruders in other solar systems. Similar comet-like intruders have already been observed in our own planetary system (see \cite{Portegies2018}). In our numerical model, former members of the Oort cloud at this large distance have a significant orbital velocity relative to the Sun of around 250--300 km s$^{-1}$. Such processes are, of course, not typical, but according to our probability study of a Solar System collision with NGC 7078, it is quite possible -- around $\sim$30\%.

In Fig. \ref{fig:dr-oort} we also emphasise the distribution of Oort cloud particles within the inner few hundred astronomical units. As can be seen, at all times during the integration process, there are always a few particles well inside 100 AU. We present a detailed discussion of the inner structure of the remaining Oort cloud system due to gravitational stripping by the $N$-body flyby cluster system in Sect. \ref{sec:disc}.

We conducted an analysis of a possible flyby of individual stars belonging to a cluster near the Sun, in which the criterion for detecting such a flyby was a relative distance of 1 pc from the Sun. We did not detect such a close flyby of the individual star. This can be explained by the possible reason for the insufficient numerical resolution of the cluster itself. For the initial distribution of stars inside the GC we used the King model with the most compact concentration parameter, King W$_0$ = 9.0 which leads to the fact that almost all the stars are within the inner 4 pc. Thus, in order to study the flybys of cluster stars near the Sun at the time of a close pass of the cluster itself, it is necessary to increase the GC particle resolution, which we plan to do in the next step of our investigation in future.

\section{Discussions}\label{sec:disc}

\begin{figure*}[htbp!]
\centering
\includegraphics[width=0.85\linewidth]{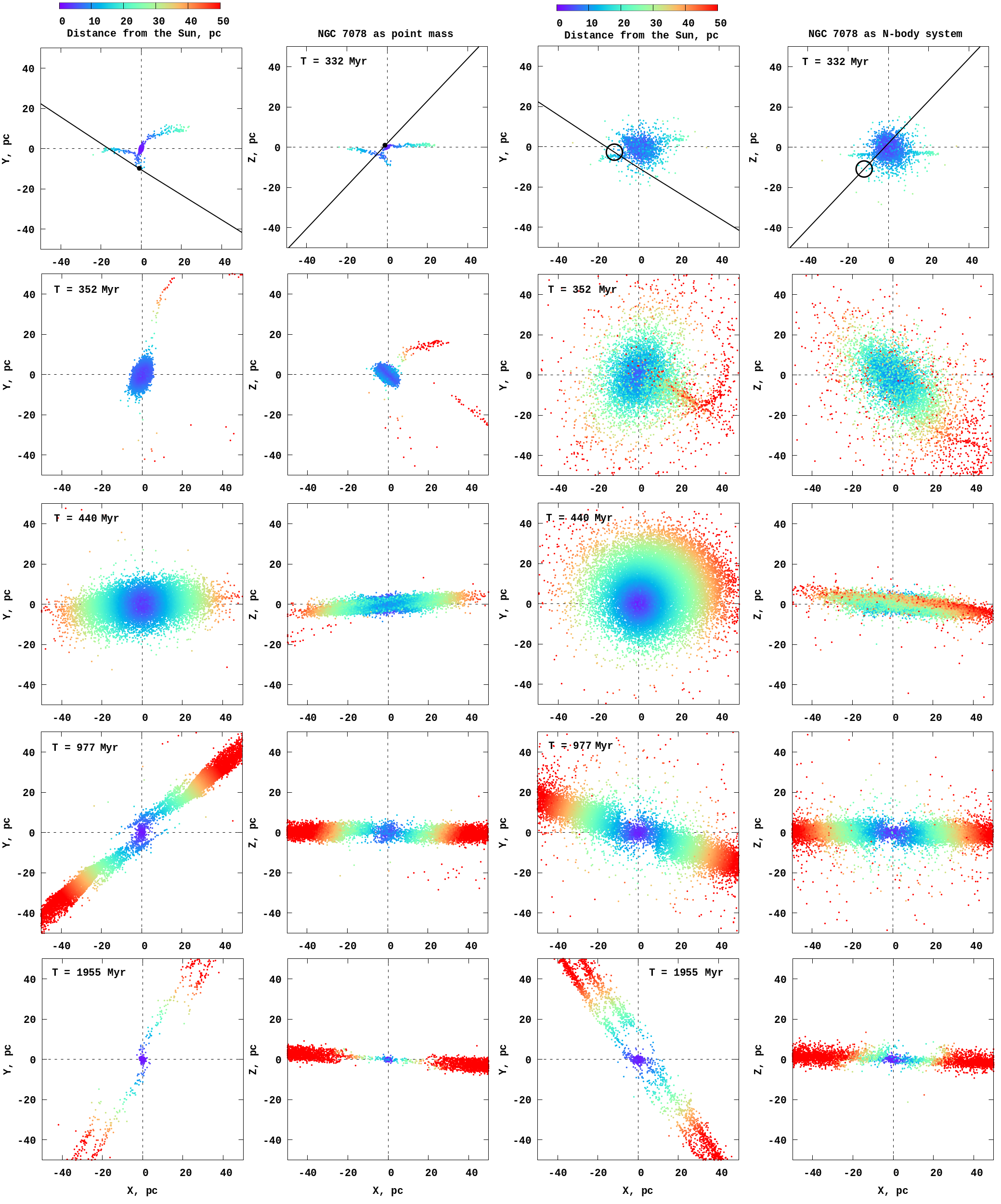}
\caption{Oort particle distribution due to the gravitational influence of the 7078 GC flyby. Two right-hand panels show the evolution of the Oort system due to the GC flyby as an $N$-body system.}
\label{fig:oort-grab-part}
\end{figure*}

\begin{figure*}[htbp!]
\centering
\includegraphics[width=0.85\linewidth]{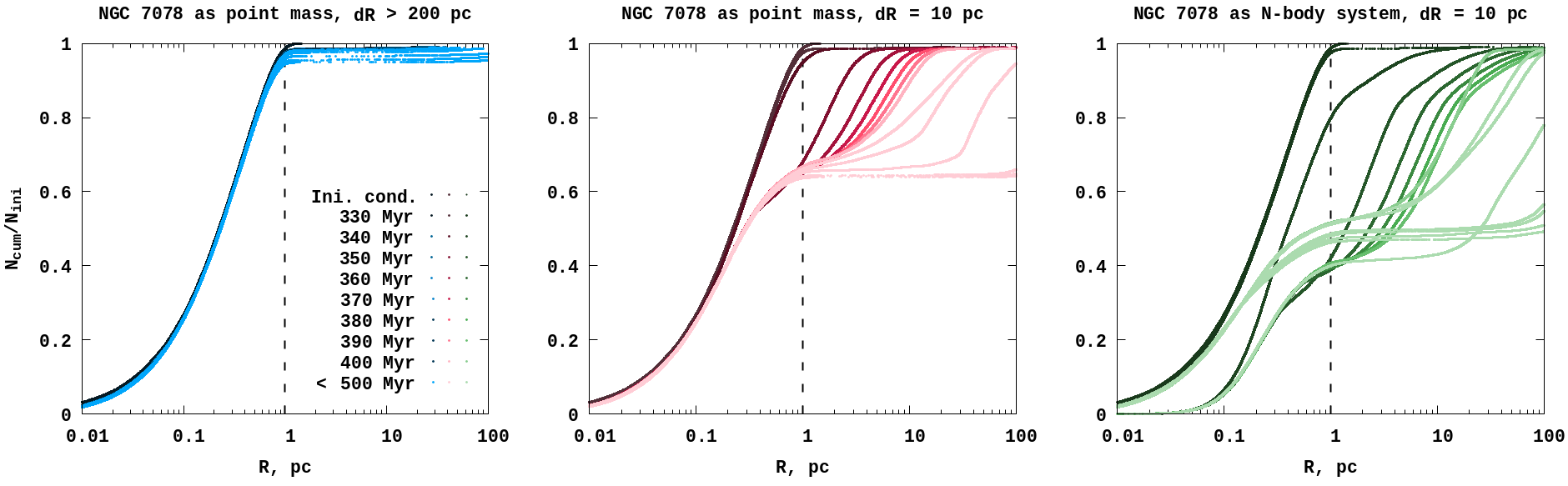}
\includegraphics[width=0.85\linewidth]{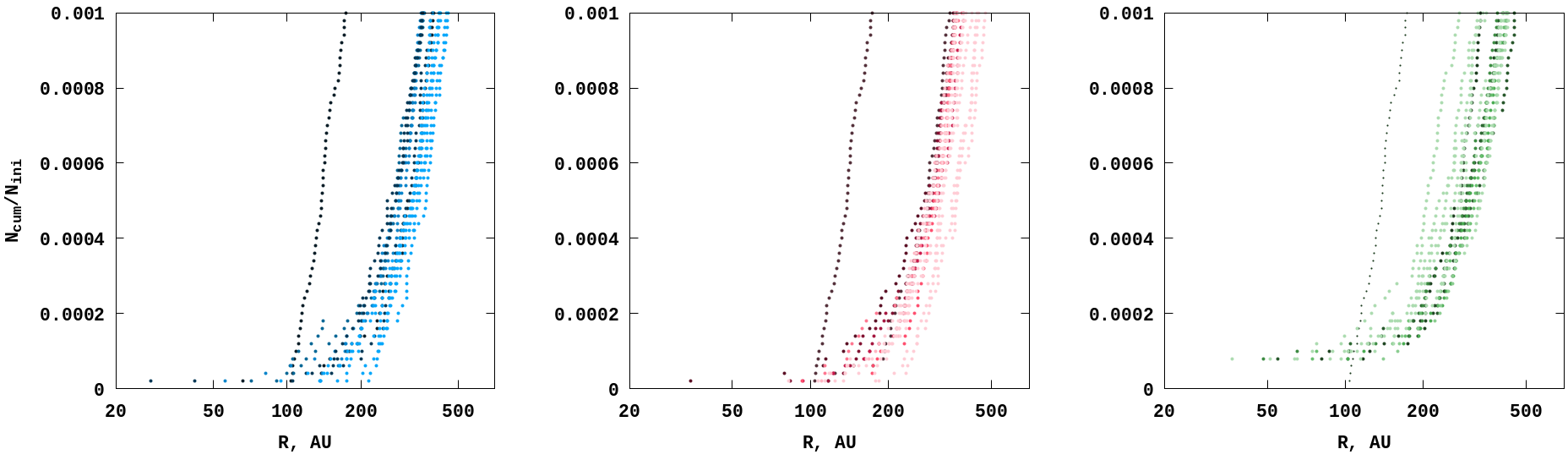}
\caption{Cumulative normalised Oort particle distributions at different times as a function of relative distance from the Sun. The left and middle panels show the particle distributions when NGC 7078 is integrated as a single physical point mass, with passages of more than 200 and 10 pc, respectively. The right panel shows the same, but with NGC 7078 integrated as an $N$-body system. The close passage in all panels occurred 332 Myr ago. The dashed black line represents the outer limit (1 pc) of the Oort particle distribution in the initial state. In the bottom panels, we present the distribution of particles in the inner Oort cloud.}
\label{fig:oort-part-distr}
\end{figure*}

The main focus of this research is to examine and analyse in detail the gravitational influence of the massive GC system (in this case, NGC 7078) on the Sun's Oort cloud system, as a result of a close flyby within a time-variable global external MW-like potential. Based on our statistical analysis, we found that, along with all other close GCs, NGC 7078 has a maximum statistical probability of a very influential close passage near our Oort cloud. To investigate this complex issue, we prepared and ran three types of simulation with different levels of complexity and detail.

In the first set of simulations, we modelled the Solar System and NGC 7078 (with 10,000 random seeds for the initial conditions of NGC 7078) as point mass particles. We inserted our particles along with their current positions and velocities (including standard deviations) into one of our MW-like time-varying external potentials (see \citep{Ishchenko2023a}). From the 5 Gyr lookback integration, we analysed the orbital crossings and identified the most promising cases for the next set of simulations (see Fig. \ref{fig:sun_orb_7078-1-gyr}).

In the next set of simulations, we shall create a more complex model of the Solar System, including the Oort cloud. During these few dozen runs, the NGC 7078 was still represented as a gravitating point mass. Based on these simulations, we identified the most suitable initial conditions of NGC 7078 versus the Oort cloud system. These conditions maximise cloud particle stripping during the close passage phase. In this case, the close passage occurred 332 Myr ago (see Figs. \ref{fig:shift} and \ref{fig:ini-pass}). The set of initial conditions also allows us to select a different fraction of unbound Oort particles after the flyby. This fraction can range widely from 10\% to 90\%.

During the final, most complex modelling stage, we re-simulated the close passing event that occurred in our dynamical system's history 332 Myr ago. To this end, we generated and analysed the long-term evolution of the full $N$-body system of NGC 7078, taking into account stellar evolution (see Figs. \ref{fig:mass-loss-7078} and \ref{fig:def-den-7078}). Ten million years prior to this event, we combined the NGC 7078 cluster with Solar System $N$-body data, incorporating the Oort cloud.

In Fig. \ref{fig:oort-grab-part} we present this specific moment of interaction between NGC 7078 and the Oort cloud system, with a separation of 10 pc between the Sun and NGC 7078. In the plot, we only present the Oort cloud particles, which are colour-coded according to their distance from the Sun. The two left-hand panels illustrate the long-term evolution of the NGC 7078 flyby as a single point mass up to 1.9 Gyr. The two right-hand panels show our system at the same moments in time, but with NGC 7078 modelled as a full $N$-body system. Contrasting the left and right plots immediately reveals the significant differences in results between simple point mass perturbations (left panels) and more complex extended GC close interactions with the cloud (right panels). As can be seen from the figure, the number of stripped particles is much more significant in the case of a distributed object containing many gravitating point masses than in the case of a simple point mass. For the live object NGC 7078, the expansion rate is significantly higher, and within the first 30 Myr after the passing event, the Oort cloud particles have spread 50 pc from the Sun. At this stage, the twisted and flattened cloud structure with extended outer tails can already be seen in the right-hand panels. As the evolution progresses, these differences become more pronounced. In both cases, at late times we clearly observe a split in the Oort cloud tail structure, indicating periodic crossings of the galactic disc during the Solar System's orbital evolution.

Even when the initial conditions of the Solar System and the NGC 7078 positions and velocities are absolutely the same, the two different set-ups (point cloud and cluster cloud) produce quite different outcomes in terms of Oort cloud stripping. These results are presented in Fig. \ref{fig:oort-part-distr}. The left panel shows the outcome of the large-distance passing (more than 200 pc) of the point mass NGC 7078 through the Oort cloud system. In this case, the maximum stripping amount is only around a few percent. The middle panel shows the same model, but with a relative close passage distance of 10 pc. In this case, the stripping rate is much higher, reaching 36\% over time. The right panel demonstrates the same set-up but with the full $N$-body system of live NGC 7078. The differences are significant, with the stripping factor reaching over 52\%. A video visualisation of the gravitational influence of NGC 7078 as an $N$-body model on the Oort cloud system can be found on YouTube\footnote{Video of the Oort cloud system's evolution due to the gravitational influence of the flyby of NGC 7078: \\~\url{https://www.youtube.com/watch?v=1_IZrjA66aA}}.

We spatially emphasise the time evolution of the inner Oort cloud structure, which is presented in the bottom panels of Fig. \ref{fig:oort-part-distr}. As can be seen, during the close passage of NGC 7078 around 330 Myr ago and for a few tens of millions of years afterwards, the number of particles inside the inner Solar System increased. We present a detailed analysis of these particles, which penetrate the inner 1 AU sphere around the Sun, in Fig. \ref{fig:ugol} (see also the discussion at the end of Sect. \ref{sec:point-oort}).

Studying the interaction between GCs such as NGC 7078 and the Oort cloud is crucial to understanding its potential impact on our planet and the Solar System. The Oort cloud, which contains comets, can be disturbed by close flybys of massive celestial objects, which can alter its structure. These disruptions may redirect comets towards the inner Solar System, potentially increasing the risk of impacts on Earth. Examining these events refines our understanding of the evolution and stability of the Solar System over time, revealing how external gravitational forces shape its dynamics.

This research also has implications for Earth's habitability. Comet activity influenced by these interactions could disrupt ecosystems or pose threats to life. Furthermore, understanding these processes enriches astrobiology by offering insights into similar dynamics that may occur in other planetary systems and shedding light on conditions that support or threaten life elsewhere. Thus, the study of interactions between GCs and systems such as the Oort cloud not only addresses questions about our planet’s safety, but also expands our knowledge of the interconnected forces shaping the Universe.

\begin{acknowledgements}
The authors thank the anonymous referee for a very constructive report and suggestions that helped significantly improve the quality of the manuscript.

Part of this work of MI was funded by the Science Committee of the Ministry of Education and Science of the Republic of Kazakhstan (Grant No. AP 22787256).

PB and MI thanks the support from the special program of the Polish Academy of Sciences and the U.S. National Academy of Sciences under the Long-term program to support Ukrainian research teams grant No.~PAN.BFB.S.BWZ.329.022.2023.

This research was supported by a grant from the European Astronomical Society thanks to the generous support of the MERAC Foundation and Springer Verlag.

\end{acknowledgements}

\bibliographystyle{mnras}  
\bibliography{part_4}   

\begin{appendix}
\onecolumn
\section{Study of the clusters orbital evolution influenced due to the measurements errors} \label{app:catal-recalc}

Table  \ref{tab:init-coord} shows a comparison of the 6D information of the GCs, focusing primarily on the heliocentric distance ($D_{\rm \odot}$), radial velocity (RV) and proper motions ($\mu_{\rm \alpha}$, $\mu_{\rm \delta}$), as estimated in the equatorial co-ordinate system. The present values are based on two versions of the catalogue \citep{VasBaum2021}, collected in 2022 and 2024 respectively. As can be seen, the differences between the two versions are quite small for most GCs, including the interesting case of NGC 7078. There are only significant differences in  $\mu_{\rm \alpha}$ for UKS 1, but this did not affect the statistical estimate of the number of close passes near the Sun (see Table \ref{tab:compar}.

\begin{table*}[hbp]
\setlength{\tabcolsep}{4pt}
\centering
\caption{GCs proper motions in the equatorial co-ordinate system at present day based on two versions of the catalogue.}
\label{tab:init-coord}
\begin{tabular}{llcccc}
\hline
\hline 
Year* & GC & $D_{\rm \odot}$  & RV & $\mu_{\rm \alpha}$ & $\mu_{\rm \delta}$ \\
 & & kpc & km /s &  mas/yr  & mas/yr \\
\hline
\hline
2022 & NGC 7078 &  10.71$\pm$0.10 & -106.84$\pm$0.30 & -0.659$\pm$0.024 & -3.803$\pm$0.024 \\
2024 & &  10.76$\pm$0.07 & -106.84$\pm$0.30 & -0.648$\pm$0.009 & -3.801$\pm$0.008 \\
2022 & UKS 1 & 15.58$\pm$0.56 & 59.38$\pm$2.63 & -2.040$\pm$0.095  & -2.754$\pm$0.063 \\
2024 & & 15.58$\pm$0.56 & 59.38$\pm$2.63 & -0.960$\pm$0.207 & -2.941$\pm$0.135 \\
2022 & BH 140 & 4.81$\pm$0.25 & 90.30$\pm$0.35 & -14.848$\pm$0.024 & 1.224$\pm$0.024 \\
2024 & & 4.81$\pm$0.25 & 90.30$\pm$0.35 & -14.901$\pm$0.069 & 1.258$\pm$0.073 \\
2022 & Djorg 1 & 9.99$\pm$0.65 & -359.18$\pm$1.65 & -4.693$\pm$0.046 & -8.468$\pm$0.041 \\
2024 & & 9.88$\pm$0.65 & -359.18$\pm$1.64 &  -4.725$\pm$0.028 & -8.470$\pm$0.021 \\
2022 & NGC 2808 & 10.06$\pm$0.11 & 103.57$\pm$0.27 & 0.994$\pm$0.024  & 0.273$\pm$0.024 \\
2024 & & 11.58$\pm$0.08 & 103.57$\pm$0.27 & 1.004$\pm$0.007 & 0.269$\pm$0.007 \\
2022 & Palomar 10 & 8.94$\pm$1.18 & -31.70$\pm$0.23 & -4.322$\pm$0.029 & -7.173$\pm$0.029  \\
2024 & & 8.94$\pm$1.18 & -31.70$\pm$0.23 & -4.326$\pm$0.010 & -7.173$\pm$0.010 \\
\hline 
\end{tabular}
\tablefoot{Years* -- data was collected in summer 2022 and 2024 from website source https://people.smp.uq.edu.au/HolgerBaumgardt/globular }
\vspace{6pt}
\end{table*}

\begin{figure*}[hbp!]
\centering
\includegraphics[width=0.99\linewidth]{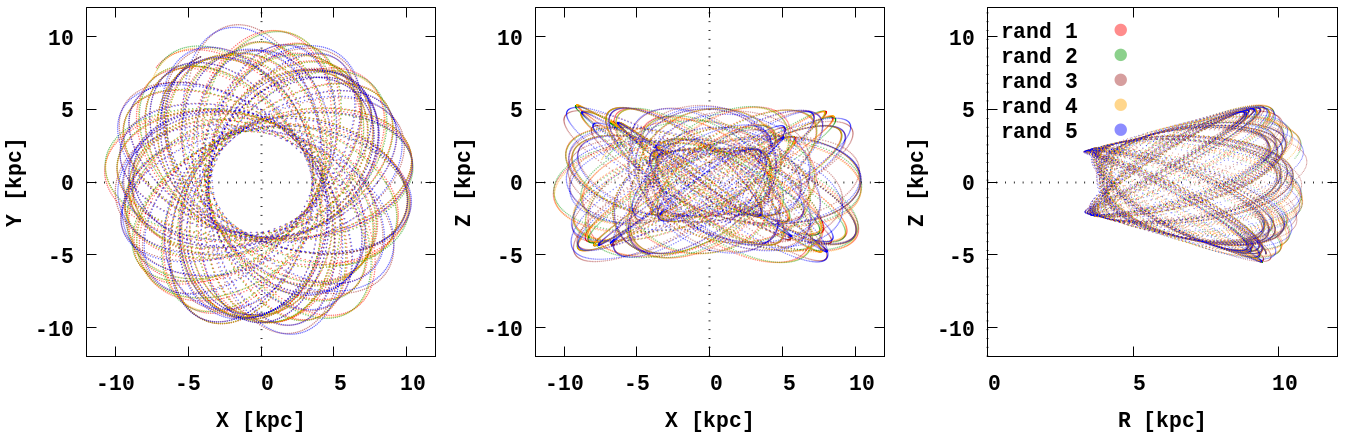}
\caption{Orbital reconstruction of the GC NGC 7078 for five different random seeds (represented by colour), with a lookback integration evolution of up to 5 Gyr in {\tt 411321} TVP external potential.}
\label{fig:orb-7078}
\end{figure*}

Fig. \ref{fig:orb-7078} illustrates how uncertainties in the initial conditions, arising from the influence of $D_{\rm \odot}$, RV and $\mu_{\rm \alpha}$, $\mu_{\rm \delta}$, can alter the shape of the orbit within the context of long-term integration (up to 5 Gyr of backward integration). As its parameters are well measured, the orbits represented by five different random seeds show minimal sensitivity to uncertainties. This leads to the assumption that the majority of NGC 7078 orbits remain stable despite variations in the initial conditions.

\begin{figure*}[hbp!]
\centering
\includegraphics[width=0.49\linewidth]{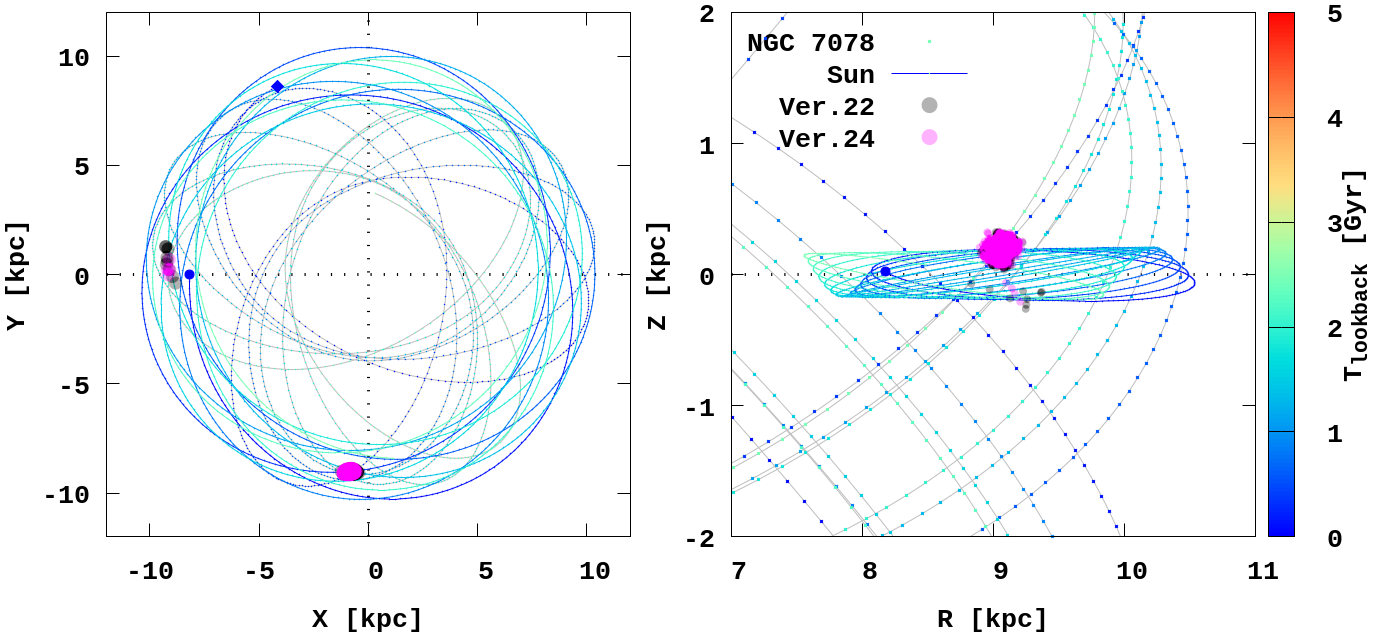}
\includegraphics[width=0.49\linewidth]{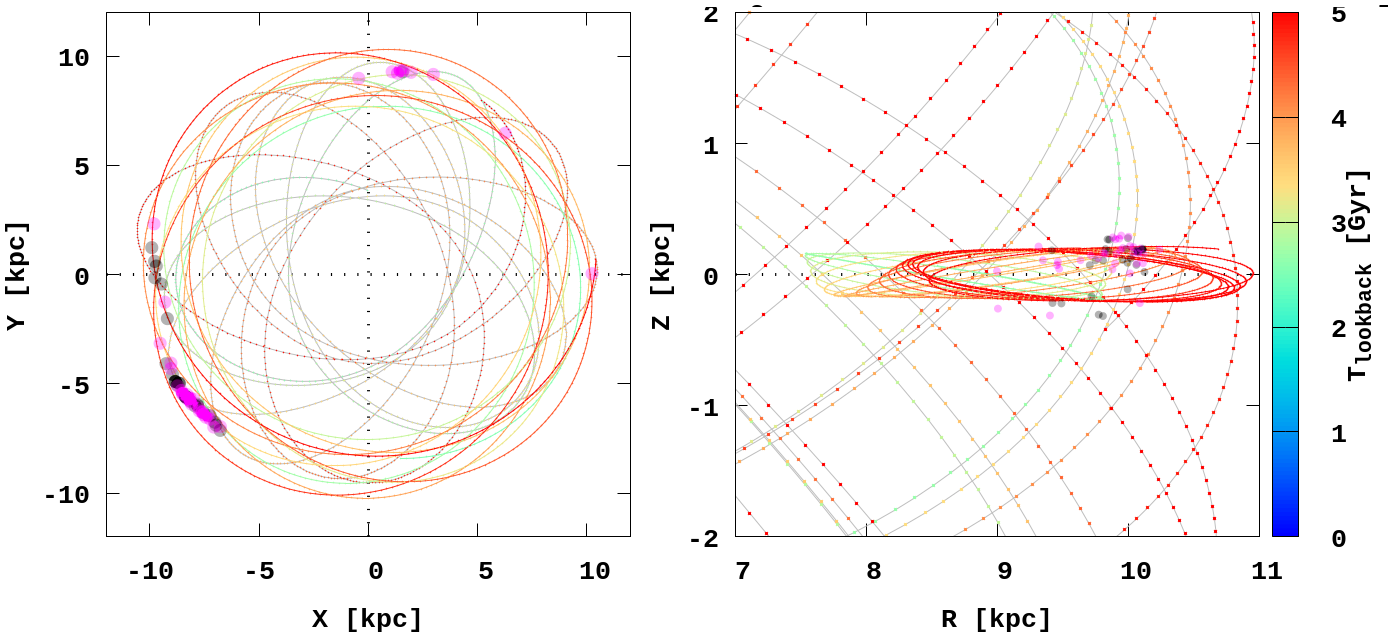}
\caption{Orbital solar evolution with a close pass of NGC 7078 during the entire integration time in {\tt 411321} TVP. The solid line represents the Sun's orbit and the coloured dots on the grey line represent the orbit of NGC 7078. The coloured circles represent collisions based on {\tt Ver.24} (magenta) and {\tt Ver.22} (black). The left panel shows the integration interval from 0 to 2.5 Gyr, and the right panel shows the interval from 2.5 to 5 Gyr in lookback time.}
\label{fig:sun_orb_7078}
\end{figure*}

In Fig. \ref{fig:sun_orb_7078} we show the orbital evolution up to 5 Gyr of integration lookback together with the Sun's orbit. The main collision is shown on the solar orbit by magenta ({\tt Ver.24}) and black ({\tt Ver.22}) dots at 332 Myr lookback time. Apart from this main close passage, we also present other collisions that occurred at: -1.7, -3.0 and -4.8 Gyr (magenta dots). But the collision probabilities of all these 3 time bins are quite low -- around 3\%.

\begin{figure}[htbp!]
\centering
\includegraphics[width=0.48\linewidth]{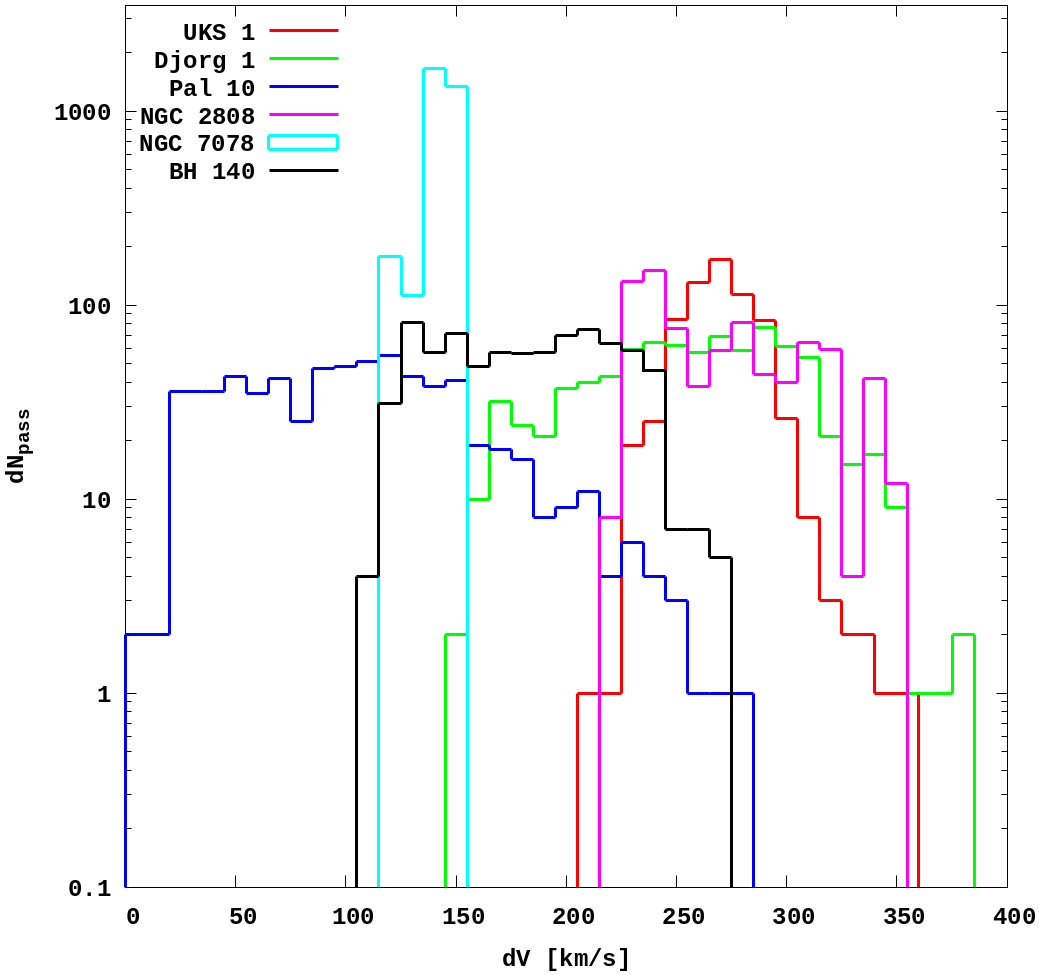}
\includegraphics[width=0.48\linewidth]{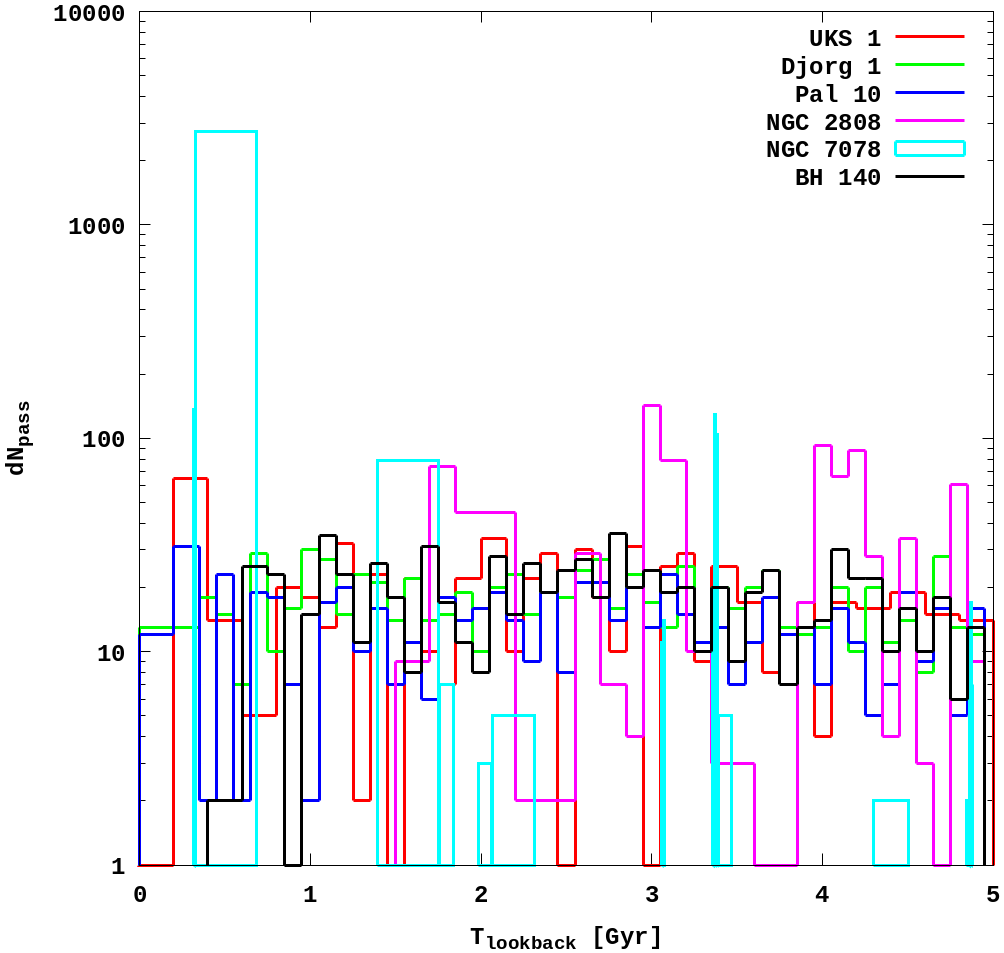}
\caption{Relative velocities (left) between GCs and Solar System and time (right) when close passes occur based on {\tt Ver.24}. Colours represent individual GCs.}
\label{fig:dr_fit2}
\end{figure}

In Fig. \ref{fig:dr_fit2} we show the relative velocities for the moments of the GCs closest passages and the corresponding time bins for our set of GCs. These statistics represent all individual random seeds for a total of 10 000 runs, with each selected GC shown in a different colour.

\end{appendix}

\end{document}